# Dynamic Animations of Journal Maps:

# Indicators of Structural Changes and Interdisciplinary Developments

Loet Leydesdorff [a] & Thomas Schank [b]

## Abstract

The dynamic analysis of structural change in the organization of the sciences requires methodologically the integration of multivariate and time-series analysis. Structural change—e.g., interdisciplinary development—is often an objective of government interventions. Recent developments in multi-dimensional scaling (MDS) enable us to distinguish the stress originating in each time-slice from the stress originating from the sequencing of time-slices, and thus to locally optimize the trade-offs between these two sources of variance in the animation. Furthermore, visualization programs like *Pajek* and *Visone* allow us to show not only the positions of the nodes, but also their relational attributes like betweenness centrality. Betweenness centrality in the vector space can be considered as an indicator of interdisciplinarity. Using this indicator, the dynamics of the citation impact environments of the journals *Cognitive Science, Social Networks*, and *Nanotechnology* are animated and assessed in terms of interdisciplinarity among the disciplines involved.

**Keywords**: interdisciplinarity, animation, change, visualization, stress, time-series, Visone.

[a] Amsterdam School of Communications Research (ASCoR), University of Amsterdam, Kloveniersburgwal 48, 1012 CX Amsterdam, The Netherlands; loet@leydesdorff.net; http://www.leydesdorff.net

[b] Technical University of Karlsruhe, Faculty of Informatics, ITI Wagner, Box 6980, 76128 Karlsruhe, Germany; schank@ira.uka.de.

## 1. Introduction

The Journal Citation Reports (JCR) of the *(Social) Science Citation Index* contain structural information about citation-relation patterns of journals at the aggregated level for each year. The aggregated journal-journal citation matrices based on this data can be analyzed in terms of their structural dimensions using, for example, factor analysis (Doreian & Farraro, 1985; Leydesdorff, 1986; Tijssen *et al*., 1987). Additionally, graph-analytical approaches enable us to visualize this data in terms of centrality measures (Freeman, 1977 and 1978/1979; De Nooy *et al*., 2005). In the case of journal maps, the clusters can be designated in terms of scientific specialties (Boyack *et al*., 2005; De Moya-Anegón *et al*., 2007).

By comparing data for different years, one can attempt to indicate structural *change* in addition to structure in the data at each moment of time. The structural model is static and contains necessarily an assumption about the number of factors (Leydesdorff, 2006). However, the number of relevant dimensions may also vary over time. If both the factor loadings and the factors themselves are allowed to vary over time, the models become unidentifiable without further assumptions. Changes in observable variation have hitherto been difficult to distinguish from changes in latent structures.

Using time-series analysis, one first has to estimate the extent to which the variation among different years is auto-correlated. If the measurements (for different moments of time) are auto-correlated, the error terms are also correlated, and this violates an assumption in a regression model. However, ARIMA—*A*uto-*R*egression *I*ntegration and *M*oving *A*verage—models (available as a routine in SPSS) have not yet been developed for a large number of variables as in the case of matrices (networks, graphs) developing over time. Furthermore, substructures of citation matrices may evolve at different speeds. For example, the cited half-life times of journals containing mainly letters or reviews are significantly different within otherwise homogenous fields of science (Leydesdorff, 2008).[3]

Recent developments in visualization and animation techniques have placed the problem of distinguishing structural change from variation once again on the agenda. Most techniques for dynamic visualizations are based on smoothing the transitions by linear interpolation between static representations in order to optimize the conservation of a mental map (Moody *et al*., 2005; De Nooy *et al*., 2005; Bender-deMoll & McFarland, 2006). In this study, we use an MDS-based algorithm to layout time series of network data dynamically by optimizing the stress both within each year and over consecutive years, that is, by optimizing in three dimensions of the data (Erten *et al*., 2004; Gansner *et al*., 2005; Baur & Schank, 2008). The new algorithm was recently implemented in *Visone*. *Visone* is a software package for the visualization of network data (Baur *et al*., 2002; Brandes & Wagner, 2004). The version with this routine added can be web-started from http://www.visone.info/dynamic or downloaded as stand-alone at http://www.leydesdorff.net/visone/index.htm.

[3] Because of the problems involved in combining multivariate analysis with time-series analysis, I turned to entropy statistics during the early 1990s. In principle, entropy statistics and its elaboration into information calculus enable us to combine static and dynamic analysis into a single design (Theil, 1972; Leydesdorff, 1991). Leydesdorff (1990) showed that prediction analysis using entropy statistics could be made more precise than predictions based on ARIMA models for the case of univariate indicators. However, the extension of entropy statistics to multivariate sets like aggregated journal-journal citation indicators did not yet allow me sufficiently to distinguish change caused by variation from structural change (Leydesdorff, 2002).

We apply the new algorithm to three evolutions of citation impact environments of journals for which an expectation about interdisciplinary developments could be specified on the basis of previous research:

1. In the case of *Cognitive Science,* we elaborate on a study by Goldstone & Leydesdorff (2006) which analyzed JCR data for 2004 in order to validate the expectation of the editors of the journal that *Cognitive Science* could play an interdisciplinary role in its relevant field (Collins, 1977). Goldstone & Leydesdorff (2006) found the highest betweenness centrality for this journal in its citation impact environment in 2004 and conjectured that betweenness centrality would remain high across the years;
2. In the case of *Social Networks*, Leydesdorff (2007) found high betweenness centrality of the journal in its citation impact environment in 2004. The field of (social) network analysis has gone through a turbulent transition period thanks to the development of internet research (Otte & Rousseau, 2002; Barabási & Albert, 1999). Fortunately, the period of this relatively recent "revolution" (Kuhn, 1962) is covered by the JCR-data (which have been available electronically since 1994). When did the interdisciplinary position of *Social Networks* emerge? Could it be sustained?
3. Nanotechnology has been a priority funding area for governments in the twenty-first century (Kostoff *et al*., 2007). Aggregated citation data for the journal

*Nanotechnology* have been available since 1996, and its position can therefore be used as an indicator of the evolution of this field during the period 1996-2006 (Leydesdorff & Zhou, 2007).

## 2. Data and Methods

The data is collected from the Journal Citation Reports of the *Science Citation Index* and the *Social Science Citation Index.* The information available in both databases was combined using relational database management. As noted, the data have been available in electronic format since 1994, but on a yearly basis. The last year available at the time of this analysis (December 2007) was 2006.

Citation matrices among journals are asymmetrical since the journals are both citing each other and cited by each other. In this study, we use only the cited structures, since we are interested in the citation impact environments. All journals citing the specific seed journal under study are included in each year. Note that the JCR does not include all journals in the tail of the distributions, but subsumes them under “All Others.” This information was not included because it cannot be made meaningful for the interpretation.

The citation patterns in the matrices are normalized using the cosine as a similarity measure (Salton & McGill, 1983; Ahlgren *et al*., 2003; Leydesdorff & Vaughan, 2006). While the Pearson correlation normalizes with reference to the arithmetic mean, the cosine normalizes with reference to the geometric mean (Jones & Furnas, 1987). This is convenient when the purpose is to visualize skewed distributions. Because the cosine varies from zero to one, a threshold is needed if one wishes to visualize structure in the data. In order to maintain consistency with previous studies (Goldstone & Leydesdorff, 2006; Leydesdorff, 2007), the threshold is set at cosine ≥ 0.2 in the first two case studies. In the case of the citation environment of *Nanotechnology*, however, the threshold had to be set at a higher value (cosine ≥ 0.5) because citation networks in the natural sciences are populated more densely than in the social sciences (Leydesdorff, 2003). Betweenness centrality is calculated on the basis of the respective vector spaces and separately for each year.

The cosine-normalized matrices are converted into a time-sliced Pajek project using *Pajek* itself and dedicated software. These files can be read into *Visone, PajekToSVGAnimation* or *SoNiA* (Social Networks Image Animator) for further processing.[4] In collaboration with the team that developed *Visone* (Görke *et al*., 2007; Baur & Schank, 2008), a dynamic solution for the animation was implemented based on optimization of stress both for each year and over the years.

The approach falls into the category of MDS-based methods. In their seminal work, Kamada and Kawai (1989) reformulated the problem of achieving graph-theoretical target distances in terms of energy optimization. They formulated the ensuing stress in the graphical representation as follows:

---

[4] *Visone* is freely available at http://visone.info, *PajekToSVGAnim* at http://vlado.fmf.uni-lj.si/pub/networks/pajek/SVGanim/default.htm , and *SoNIA* at http://www.stanford.edu/group/sonia/ .

$$S = \sum_{i \neq j} s_{ij} \qquad \text{with} \qquad s_{ij} = \frac{1}{d_{ij}^2}(\|x_i - x_j\| - d_{ij})^2 \qquad (1)$$

where for each pair of nodes *i* and *j,* the parameter $d_{ij}$ is the distance making the shortest path between this pair. However, Kruskal's (1973) stress value for MDS was defined differently (e.g., Kruskal & Wish, 1978; Borgatti, 1998), notably as follows:

$$S = \sqrt{\frac{\sum_{i \neq j} (\|x_i - x_j\| - d_{ij})^2}{\sum_{i \neq j} d_{ij}^2}} \qquad (2)$$

Equation 2 differs from Equation 1 by taking the square root and because of the weighing of each term with $1/d_{ij}^2$ in the numerator of Equation 1. This weight is crucial for the quality of the layout, but defies normalization with $\sum d_{ij}^2$ in the denominator and hence the comparability between these two stress values.[5]

Gansner *et al.* (2005) improved on the algorithm of Kamada & Kawai (1989) by minimizing a so-called majorant of the stress-function *S*. Using a number of empirical

[5] One could consider developing a stress measure that is normalized, but based on Kamada & Kawai (1989) using the following normalization:

$$S = \sqrt{\frac{\sum_{i \neq j} \frac{1}{d_{ij}^2}(\|x_i - x_j\| - d_{ij})^2}{\sum_{i \neq j} 1}}$$

We intend to develop this methodological argument systematically in a next study, for both static and dynamic cases (to be discussed below; Leydesdorff *et al.*, in preparation).

cases, they showed that their approach converges much faster, is less sensitive to local minima, and further minimizes the stress function provided in Equation 1. In addition to these methodological advantages, the majorant can also be implemented using an algorithm that is more compact and faster than that of Kamada & Kawai (1989).

We extended this algorithm to layout dynamic networks (Baur & Schank, 2008). The corresponding dynamic stress function is provided by the following equation:

$$S = \left[ \sum_t \sum_{i \neq j} \frac{1}{d_{ij,t}^{\;2}} (\| x_{i,t} - x_{j,t} \| - d_{ij,t})^2 \right] + \left[ \sum_{1 \leq t < |T|} \sum_i \omega \, \| x_{i,t} - x_{i,t+1} \|^2 \right] \quad (3)$$

In Equation 3, the left-hand term is equal to the static stress, while the right-hand term adds the dynamic component, namely the stress between subsequent years. If the weighting factor ω for this dynamic extension is set equal to zero, the method is equivalent to the static analysis and the layout of each time frame is optimized independently. The dynamic extension penalizes drastic movements of the position of node *i* at time *t* ($\vec{x}_{i,t}$) toward its next position ($\vec{x}_{i,t+1}$) by increasing the stress value. Thus, stability is provided in order to preserve the mental map between consecutive layouts so that an observer can easily identify corresponding graph structures. Preserving the mental map is a crucial point in computing dynamic layouts (Misue *et al.*, 1995).

In other words, the configuration for each year can be optimized in terms of the stress in relation to the solutions for previous years and in anticipation of the solutions for following year. In principle, the algorithm allows us (and *Visone* enables us) to extend this to more than a single year, but in this study the optimization is extended by only one year in both directions (that is, including $t + 1$ and $t - 1$). Note that this approach is different from the approach that takes the solution for the previous moment in time as a starting position for iterative optimization according to Equation 1. The nodes are not repositioned given a previous configuration, but the previous and the next configurations are included in the algorithmic analysis for each year.

Technically, the equation to be optimized computes iteratively a new position for each node ($x_i$) on dimension $d$, as follows:

$$\text{new}-x_{i,t}^{(d)} = \frac{\left[\sum_{j \neq i} w_{ij,t}\left(x_{j,t}^{(d)} + d_{ij,t}\frac{x_{i,t}^{(d)} - x_{j,t}^{(d)}}{\left\| x_{i,t} - x_{j,t} \right\|}\right)\right] + \omega\ \left(x_{i,t-1}^{(d)} + x_{i,t+1}^{(d)}\right)}{\left[\sum_{j \neq i} w_{ij,t}\right] + 2\omega} \tag{4}$$

until the aggregated stress comes below a threshold value or during a fixed number of iterations. Again, the left-hand term (between brackets in both the numerator and the denominator of Equation 4) accounts for the static solution, while the right-hand terms contain the extensions with the stress in comparison to the previous ($t$–1) and next ($t$+1) moments in time. Higher values of the weighting factor for the dynamic extension (ω) result in increased stability of the representations over the years.

Like various other parameters, one can experiment with this weight function using *Visone*. *Visone* offers as a further advantage that one can animate using the sizes of the nodes as indicators of the various centrality measures. Other animation programs like *PajekToSVGAnim* or *SoNIA* cannot accommodate these values in the animations without extensive preprogramming. *Pajek*, for example, stores this information separately in a vector. In order to enhance the readability of the animations, isolates and small components which were not related to the largest component in any of the years under study, were removed.

## 3. Results

### *3.1 Cognitive Science*

Goldstone & Leydesdorff (2006) analyzed the citation impact environment of *Cognitive Science* in 2004. The journal had 1,113 citations in this year, spread across 180 journals; 164 of these journals formed a single component when a threshold of cosine ≥ 0.2 was applied (Figure 1). The two largest clusters of journals citing *Cognitive Science* in 2004 were cognitive psychology and computer science, with neuroscience rather densely interconnected with psychology. Education, linguistics, and philosophy journals all had presences in the cited environments of the journal that are stronger than in its citing networks.

In other words, *Cognitive Science* was cited by journals in otherwise poorly related fields. This can be seen upon visual inspection of Figure 1: betweenness centrality is used as a measure for the sizes of the nodes. The betweeness centrality of *Cognitive Science* in this vector space was 30.3%, whereas the second-largest betweeness value in 2004 was 5.1% for *Annual Review of Psychology*.

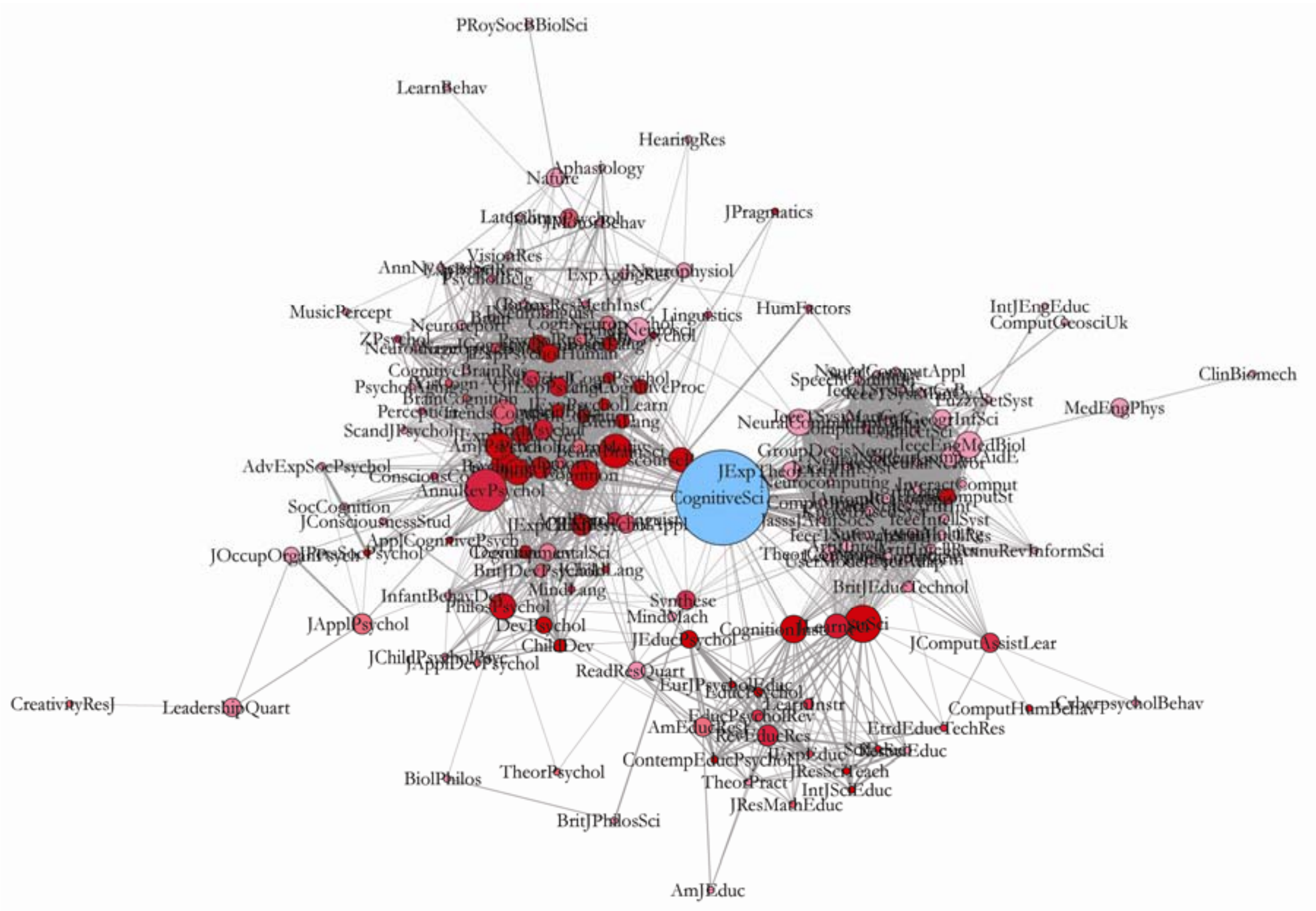

**Figure 1**: Betweenness centrality of 164 journals in the citation impact environment of *Cognitive Science* in 2004;  cosine > 0.2.

The pronounced betweenness centrality of the journal accords with the aspiration of the editors to provide an interdisciplinarity meeting place for scholars approaching problems of cognition from different disciplinary angles (Collins, 1977). Goldstone & Leydesdorff (2006, at p. 988) explored this property longitudinally on the basis of a comparison with similar data for 1988, and found also in that year a strongly mediating role for the journal among psychology, computer science, and education.

Using exactly the same methods as these authors, an animation of the journal's betweenness centrality in the vector space during the period 1994-2006 is available at http://www.leydesdorff.net/journals/cognsci/index.htm. The animation shows that the betweenness centrality of the journal in this network is high in all these years, but that the relevant environments change over the years, with the exception of the educational field, which is connected to the field of cognitive psychology by the citation patterns of the journals *Cognitive Science* and *Cognitive Instruction*, respectively. However, the relation to journals in the computer sciences that was found in 2004 can be considered as an exception rather than the rule.

In summary, *Cognitive Science* functions at the margins of cognitive psychology in virtually all the years with the specific function of relating this specialty to specialties in other (sub)disciplines, that is, across disciplinary divides. Its citation relation with education science research is stable. However, the original vision of Collins (1977) of starting the journal in order to add "another trapping in the formation of a new discipline" (p. 1) has not been realized. The journal has a strongly interdisciplinary character, but the mother disciplines (psychology and education research) have remained dominant frames of reference for the journal's "interdisciplinary" development.

### *b. Social Networks*

When *Social Networks* was launched in 1978, it added a new communication channel to a set of journals that focus on quantitative methodologies in sociology. The journal can also be considered as a sociological pendant of journals about psychometrics, econometrics, biometrics, etc. However, models of neural networks in psychology and biology focus on dynamic properties of networks, while social network analysis was developed primarily as a toolbox for the analysis of the structural properties of networks (Burt, 1982 and 1987; Freeman, 1977 and 1978/1979; Wasserman & Faust, 1994). The development of the scholarly journal and the corresponding specialty have been closely linked to the development of computer programs like *UCINet*, *Structure*, *Mage*, *Visone*, and *Pajek* (Freeman, 2004).

The advent and growth of the Internet during the 1990s generated huge interest in the techniques developed within this field (e.g., centrality measures) among both social and natural scientists. Figure 2 shows the resulting configuration in terms of the betweenness centrality of the journal in 2004: 43 of the 54 journals citing *Social Networks* during this year are connected at the level of cosine ≥ 0.2 (Leydesdorff, 2007). The journal has a central position in Figure 2 between the fields of sociology and management science, on the one hand, and branches of the exact sciences focusing on network analysis, on the other. The *Journal of Mathematical Sociology* and *Organization Science* support this construction of an interface between clusters with social and natural science journals, respectively.

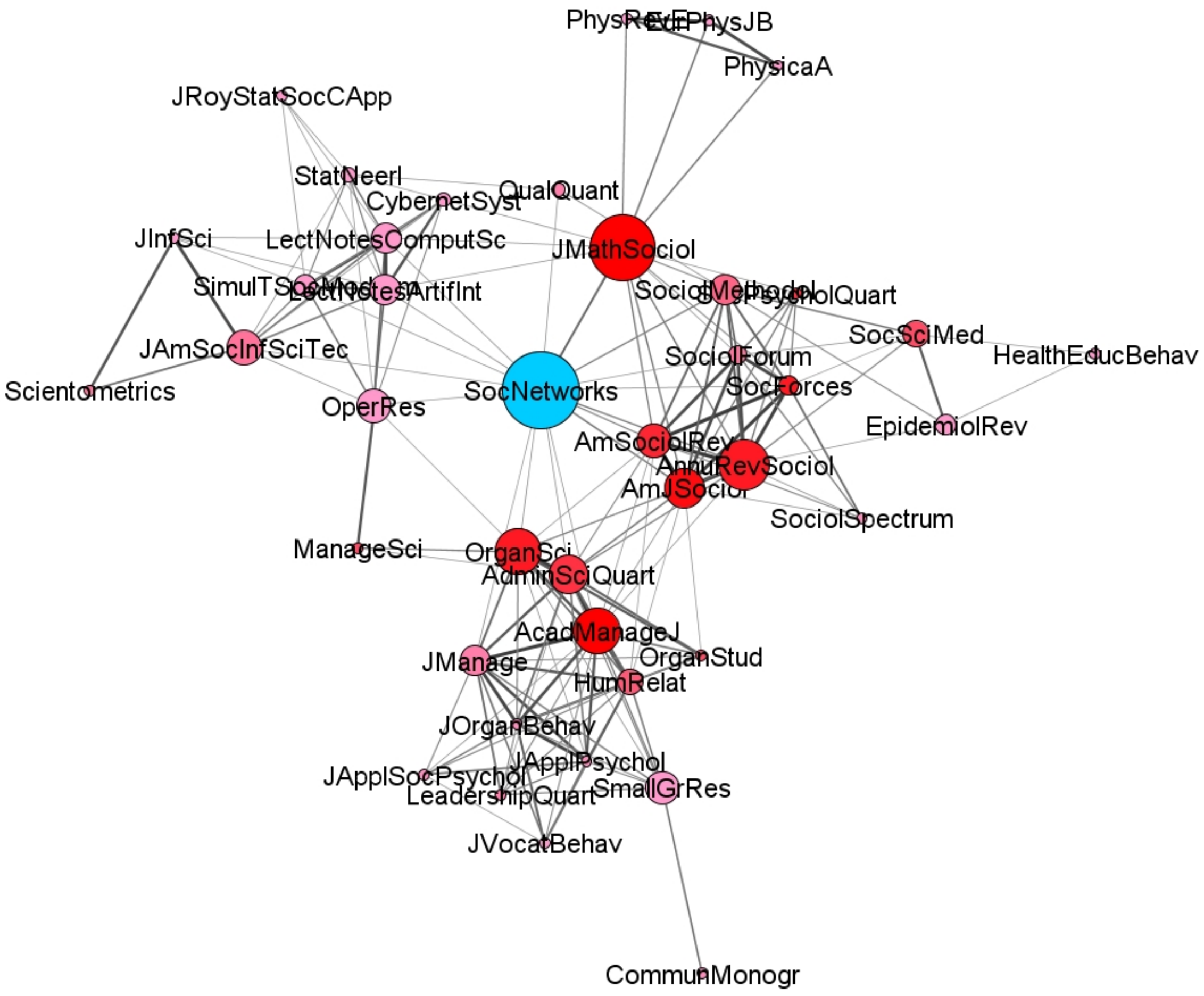

**Figure 2**: Betweenness centrality of 43 journals in the vector space of the citation impact environment of *Social Networks* (cosine ≥ 0.2).

Can this interdisciplinary position be maintained over the years? The animation for the period 1994-2006 (available at http://www.leydesdorff.net/journals/socnetw/index.htm) suggests differently. In the first years, *Social Networks* is clearly visible as a sociology journal—related most closely to the *American Sociological Review* and the *American Journal of Sociology* as leading journals in the field—and in all years the citation pattern of the journal is primarily attributed to this group. However, the journal definitively gained interest in a set of management journals during the second half of the 1990s.

In some years more than others, the journal's citation pattern moved away from the sociology core group, but this movement was not consolidated. Thus, the journal's interdisciplinary ambitions are offset against its disciplinary identity. This leads to a pattern of alteration between years in which the disciplinary citation pattern competes with the interdisciplinary orientation. As noted, the position of the journal between sociology and management science was consolidated during the last decade.

*c. Nanotechnology*

The journal *Nanotechnology* was included in the *Science Citation Index* in 1996, and thus the animation (available at http://www.leydesdorff.net/journals/nanotech/index.htm) covers only the decade 1996-2006. The animation shows that this journal was first embedded in the field of "applied physics" journals, but then became an increasingly central focus of attention within this specialty towards the end of the millennium. In the period 2000-2003, nanotechnology became a priority funding area in most advanced nations (Zhou & Leydesdorff, 2006; Kostoff *et al*., 2007). At the level of aggregated journal-journal citations, this "revolution"—in the funding?—led to a reorganization of the interface between applied physics and physical chemistry.

The journal *Nanotechnology* played an important role in this reorganization of interdisciplinary development at the field level. First, the attention of citing journals in the field of applied physics was focused on this journal. Thereafter, chemistry journals began to pay increasingly attention to this field. In 2001, *Nanotechnology* as a specialist journal took the interdisciplinary role at this interface over from *Science* which had made this connection in 2000 (Figure 3).

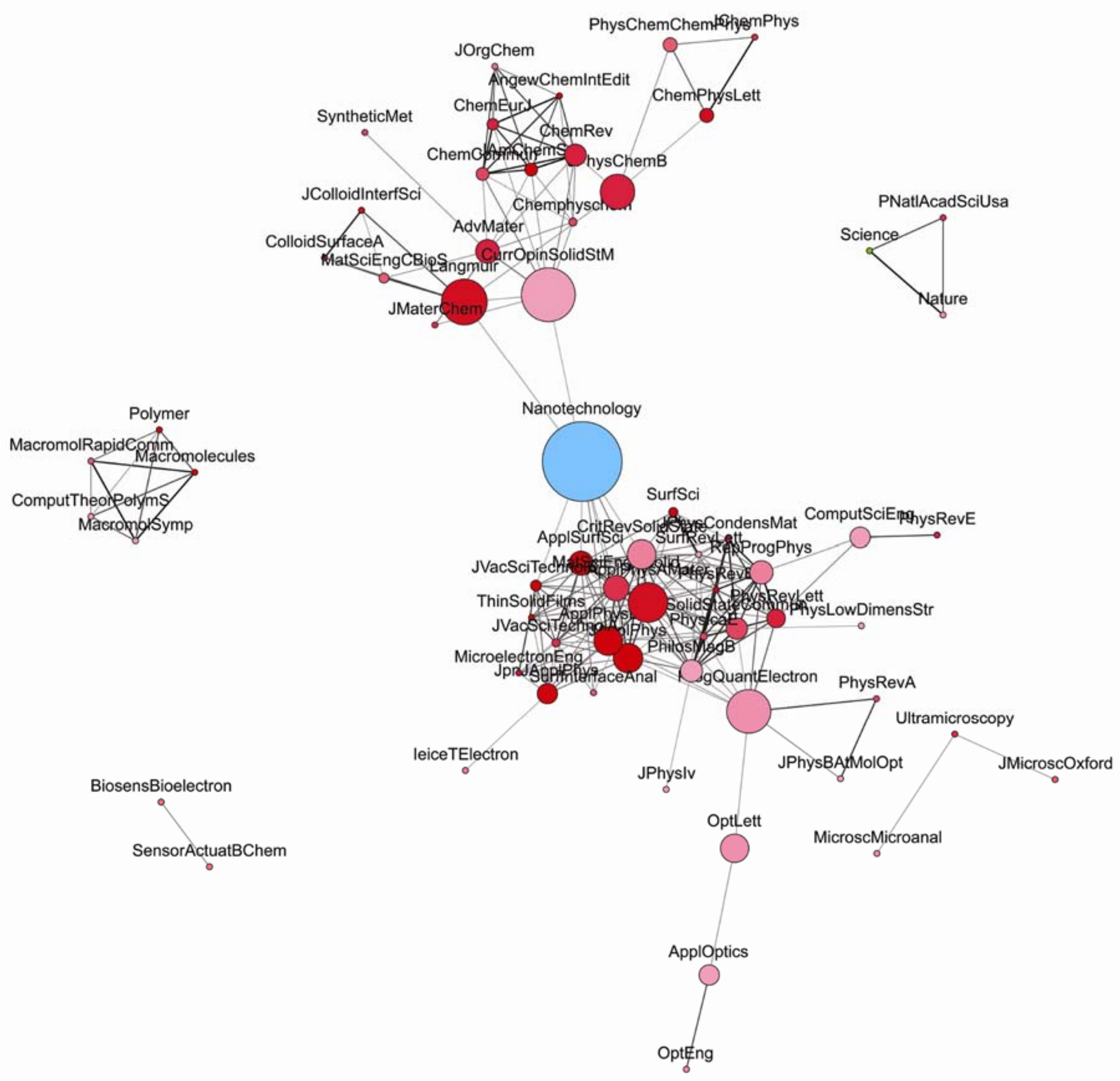

**Figure 3**: Betweenness centrality of *Nanotechnology* between clusters of journals in applied physics and chemistry, in 2001.

New journals emerged in the years thereafter, among them *Nano Letters* published by the influential American Chemical Society since 2001. As could be expected (Bensman, 1996), this latter journal took the lead in terms of the impact factors among the specialist journals at the interface between applied physics and physical chemistry. At the same time, the multidisciplinary journal *Science* began to participate in the fine-grained citation environment of these specialisms, and the journal *Nanotechnology* lost its catalyzing function at the interface.

Figure 4 shows the catalyzing role of the journal during the transition period in terms of the development of its betweenness centrality in the field. Before and after the transition the journal was firmly embedded and did not play a role at an interdisciplinary interface. During the period of transition, however, the journal reflected the turmoil in its citation environment. Betweenness centrality in its vector space increased to 27.5% in 2001. After the transition period this interdisciplinary role was taken over by a number of nanoscience journals that had emerged in the meantime (Leydesdorff & Zhou, 2007; Leydesdorff, forthcoming). The journal *Science* (and to a lesser extent *Nature*) played a role in shaping the interface in 2000 and thereafter.

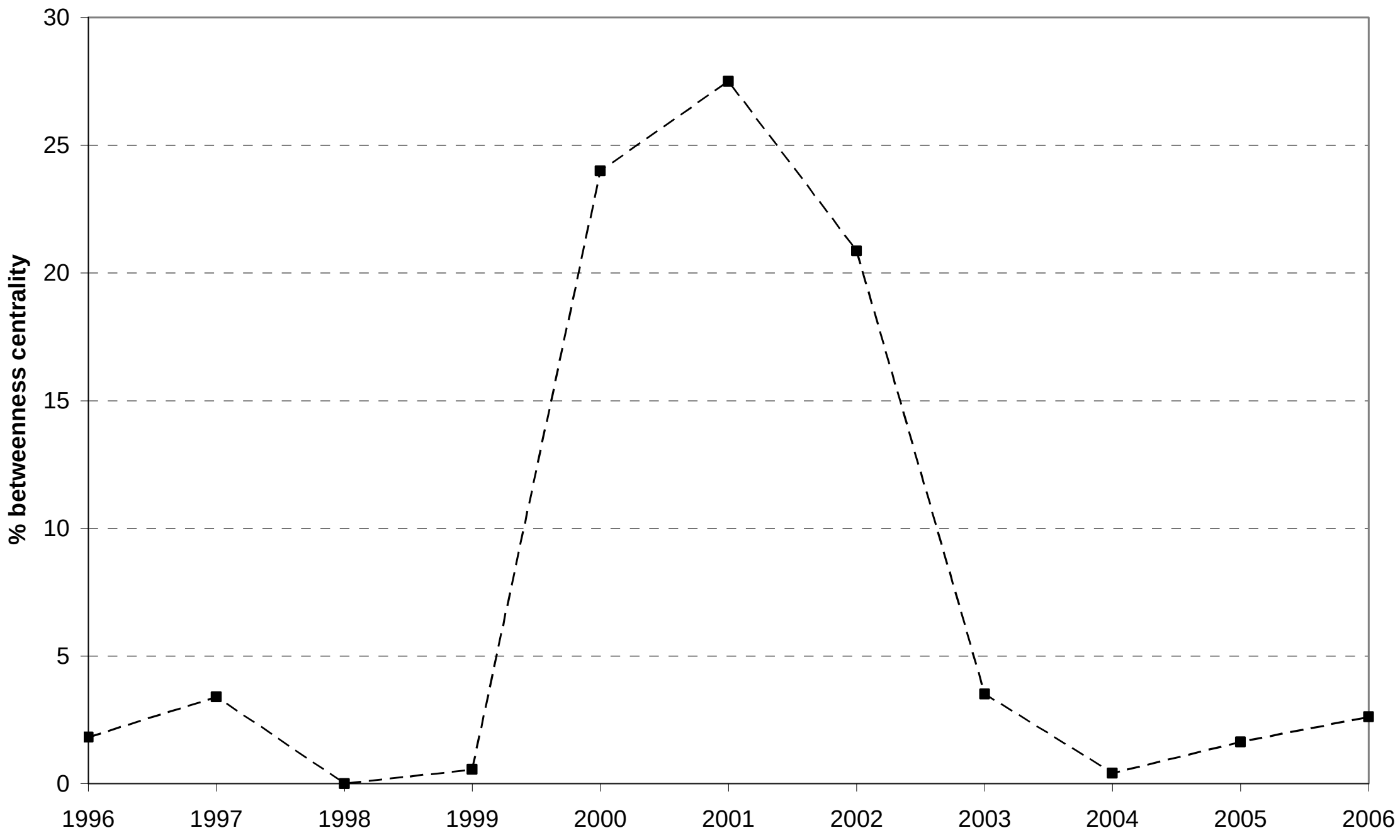

**Figure 4**: The development of betweenness centrality of *Nanotechnology* in its citation impact environment during the period 1996-2006.

## 4. Conclusions and discussion

The interdisciplinarity of new developments in science has a high policy relevance. Proponents of new developments often proclaim that the existing disciplinary structures do not sufficiently honor the potential benefits of intellectual synergy in interdisciplinary projects. Policy-makers are sympathetic to these claims, since integration and problem-orientation is emphasized as opposed to differentiation and specialization.

In Leydesdorff *et al.* (1994), we argued that new developments in science first manifest as specific journals focusing on the issues under study. The new journal attracts the attention of scholars in neighboring disciplines, and this can be measured in terms of being cited. (In terms of probabilistic entropy, one might say that the new developments increase the local temperature in the database [Kostoff, 1997; Leydesdorff, 2003].) The focus of this study was to examine whether this specific function of interdisciplinary mediation among disciplinary structures could also be stabilized over time. The case studies teach us that this is not always the case, and such mediation seldom leads to the stabilization of an interface beyond *two* specialties (Leydesdorff, 1992).

*Cognitive Science* provides an example of a journal that deliberately searches its relevant environments for potential audiences for new knowledge claims that are obtained mainly from new developments in cognitive psychology. The interface with education research was firmly stabilized during this decade, and the journal receives a sufficiently large number of citations from outside its original discipline to maintain high betweenness centrality in its relevant citation environment. However, the relations with cognitively more remote specialties remain in flux.

The journal *Social Networks* first crystallized as a methods journal in sociology during the 1980s, but then experienced the Internet revolution during the 1990s. Actually, the journal could have been expected to develop into a center for these activities, but this did not happen. The relationship with management journals was firmly established on the side of the social sciences involved, but the physics journals involved in Internet research (e.g., *Physics Review E*) do not cite this journal regularly enough for it to become an interdisciplinary node between the natural and social sciences. Instead, the citation pattern in most of the years has remained encapsulated in the mother discipline.

In the case of nanotechnology, the intellectual fields of applied physics and relevant chemistry have undergone reorganization during the period under study. The field of nanotechnology emerged as a specific domain of application in physics, on the one hand, and in chemistry, on the other, with a specific focus on nanostructures (e.g., fullerenes and nanotubes; Lucio-Arias & Leydesdorff, 2007). The journal *Nanotechnology,* which existed before the "revolution"—or the "surge in funding"— played a role among other journals in applied physics first by catching the attention to the nano-field among physicists and then in forging a relationship with chemistry. The latter relationship transformed the field of advanced material sciences into a citation cluster at the interface of applied physics and physical chemistry.

In summary, the claim of "interdisciplinarity" eventually seems in practice to lead to the emergence of a specific interface between two existing specialties and the potential reorganization of that interface into a coevolution. This accords with the evolutionary expectation, because the interfacing of more than two disciplinary codes of communication could easily lead to confusion and thus impede intellectual development. The reach beyond a single interface tends to remain incidental, and countervailing tendencies towards intellectual differentiation and disciplinary identification can also be expected. These different tendencies may even lead to alterations over the years.

Furthermore, we did not witness the emergence of stable interfaces between the social and natural sciences in these case studies, although the development of such an interface could have been expected in the cases of *Social Networks* and *Cognitive Science*. Perhaps the formation and stabilization of an interdisciplinary interface between the social and natural sciences would be “a bridge too far” given the centrifugal forces of cognitive codes of communication (e.g., the use of very different methodologies) in each of the disciplines involved.

## Acknowledgements

The authors are grateful to Diana Lucio-Arias, Wouter de Nooy, and Andrea Scharnhorst for the discussion of previous drafts.

## References

Ahlgren, P., Jarneving, B., & Rousseau, R. (2003). Requirement for a Cocitation Similarity Measure, with Special Reference to Pearson's Correlation Coefficient. *Journal of the American Society for Information Science and Technology,* 54(6), 550-560.

Barabási, A. L., & Albert, R. (1999). Emergence of Scaling in Random Networks. *Science,* 286(5439), 509.

Baur, M., Benkert, M., Brandes, U., Cornelsen, S., Gaertler, M., Köpf, B., Lerner, J., & Wagner, D. (2002). Visone Software for Visual Social Network Analysis. Proc. $9^{th}$ Intl. Symp. Graph Drawing (GD’01). *Lecture Notes of Computer Science,* 2265, 554-557.

Baur, M. & Schank, T., 2008. *Dynamic Graph Drawing in Visone*. Technical University Karlsruhe, Karlsruhe. Available at http://i11www.ilkd.uni-karlsruhe.de/people/schank/publications/bs-dgdv-08.pdf (retrieved on Apr. 22, 2008).

Bender-deMoll, S., & McFarland, D. A. (2006). The art and science of dynamic network visualization. *Journal of Social Structure*, 7(2).

Bensman, S. J. (1996). The structure of the library market for scientific journals: The case of chemistry. *Library Resources & Technical Services,* 40, 145-170.

Borgatti, S. (1998). Social Network Analysis Instructional Website, at http://www.analytictech.com/networks/mds.htm (retrieved on Jan. 25, 2008).

Boyack, K. W., Klavans, R., & Börner, K. (2005). Mapping the Backbone of Science. *Scientometrics,* 64(3), 351-374.

Brandes, U., & Wagner, D. (2004). visone – Analysis and Visualization of Social Networks. In M. Jünger & P. Mtuzel (Eds.), *Graph Drawing Software* (pp. 321-340): Springer.

Burt, R. S. (1982). *Toward a Structural Theory of Action*. New York, etc.: Academic Press.

Burt, R. S. (1987). *Structure. Version 3.2, Technical Report # TR2.* New York: Columbia University.

Collins, A. (1977). Why cognitive science? *Cognitive Science,* 1, 1-2.

De Moya-Anegón, F., Vargas-Quesada, B., Chinchilla-Rodríguez, Z., Corera-Álvarez, E., Munoz-Fernández, F. J., & Herrero-Solana, V. (2007). Visualizing the marrow of science. *Journal of the American Society for Information Science and Technology,* 58(14), 2167-2179.

De Nooy, W., Mrvar, A., & Batagelj, V. (2005). *Exploratory Social Network Analysis with Pajek*. New York: Cambridge University Press.

Doreian, P. (1986). A Revised Measure of Standing of Journals in Stratified Networks,. *Scientometrics 11*, 63-72.

Doreian, P., & Fararo, T. J. (1985). Structural Equivalence in a Journal Network. *Journal of the American Society for Information Science,* 36, 28-37.

Erten, C., Harding, Ph. J., Kobourov, S. G., Wampler, K., & Yee, G. V. (2004). GraphAEL: Graph animations with evolving layouts. Pages 98-110 in Liotta, G., Editor, *Graph Drawing*, Perugia, Italy, September 21-24, 2003; Springer, 2004.

Freeman, L. C. (1977). A Set of Measures of Centrality Based on Betweenness. *Sociometry,* 40(1), 35-41.

Freeman, L. C. (1978/1979). Centrality in Social Networks: Conceptual Clarification. *Social Networks,* 1, 215-239.

Freeman, L. C. (2004). *The Development of Social Network Analysis: A Study in the Sociology of Science*: BookSurge.

Gansner, E. R., Koren, Y., & North, S. (2005). Graph Drawing by Stress Majorization. In J. Pach (Ed.), *Graph Drawing, Lecture Notes in Computer Science* (Vol. 3383, pp. 239-250). Berlin/Heidelberg: Springer.

Goldstone, R., & Leydesdorff, L. (2006). The Import and Export of *Cognitive Science*. *Cognitive Science,* 30(6), 983-993.

Görke, R., Schank, T., & Wagner, D. (2007). Static and Dynamic Visual Analysis of a Co-Author Network. Available at http://i11www.iti.uni-karlsruhe.de/people/schank/gd07cont/schank-gd07cont.pdf (30 January 2008).

Jones, W. P., & Furnas, G. W. (1987). Pictures of Relevance: A Geometric Analysis of Similarity Measures. *Journal of the American Society for Information Science,* 36(6), 420-442.

Kamada, T., & Kawai, S. (1989). An algorithm for drawing general undirected graphs. *Information Processing Letters,* 31(1), 7-15.

Kostoff, R. N., Eberhart, H. J., & Toothman, D. R. (1997). Database tomography for information retrieval. *Journal of Information Science,* 23(4), 301.

Kostoff, R. N., Koytcheff, R. G., & Lau, C. G. Y. (2007). Global nanotechnology research metrics. *Scientometrics,* 70(3), 565-601.

Kruskal, J. B., & Wish, M. (1978). *Multidimensional Scaling*. Beverly Hills, CA: Sage Publications.

Kuhn, T. S. (1962). *The Structure of Scientific Revolutions*. Chicago: University of Chicago Press.

Leydesdorff, L. (1986). The Development of Frames of References. *Scientometrics* 9, 103-125.

Leydesdorff, L. (1990). The Prediction of Science Indicators Using Information Theory. *Scientometrics 19*, 297-324.

Leydesdorff, L. (1991). The Static and Dynamic Analysis of Network Data Using Information Theory. *Social Networks,* 13, 301-345.

Leydesdorff, L. (1992). Irreversibilities in Science and Technology Networks: An Empirical and Analytical Approach. *Scientometrics,* 24, 321-357.

Leydesdorff, L. (2002). Indicators of Structural Change in the Dynamics of Science: Entropy Statistics of the *SCI Journal Citation Reports*. *Scientometrics,* 53(1), 131-159.

Leydesdorff, L. (2003). Can Networks of Journal-Journal Citations Be Used as Indicators of Change in the Social Sciences? *Journal of Documentation,* 59(1), 84-104.

Leydesdorff, L. (2006). Can Scientific Journals be Classified in Terms of Aggregated Journal-Journal Citation Relations using the Journal Citation Reports? *Journal of the American Society for Information Science & Technology,* 57(5), 601-613.

Leydesdorff, L. (2007). "Betweenness Centrality" as an Indicator of the "Interdisciplinarity" of Scientific Journals. *Journal of the American Society for Information Science and Technology,* 58(9), 1303-1309.

Leydesdorff, L. (2008a). *Caveats* for the Use of Citation Indicators in Research and Journal Evaluation. *Journal of the American Society for Information Science and Technology,* 59(2), 278-287.

Leydesdorff, L. (2008b). The delineation of nanoscience and nanotechnology in terms of journals and patents: a most recent update. *Scientometrics* 76(1), forthcoming.

Leydesdorff, L., Cozzens, S. E., & Besselaar, P. v. d. (1994). Tracking Areas of Strategic Importance using Scientometric Journal Mappings. *Research Policy,* 23, 217-229.

Leydesdorff, L., Schank, T., Scharnhorst, A., & De Nooy, W. (in preparation). Animating the Development of *Social Networks* over Time using a Dynamic Extension of Multidimensional Scaling.

Leydesdorff, L., & Vaughan, L. (2006). Co-occurrence Matrices and their Applications in Information Science: Extending ACA to the Web Environment. *Journal of the American Society for Information Science and Technology,* 57(12), 1616-1628.

Leydesdorff, L., & Zhou, P. (2007). Nanotechnology as a Field of Science: Its Delineation in Terms of Journals and Patents. *Scientometrics,* 70(3), 693-713.

Lucio-Arias, D., & Leydesdorff, L. (2007). Knowledge emergence in scientific communication: from "fullerenes" to "nanotubes". *Scientometrics,* 70(3), 603-632.

Misue, K., Eades, P., Lai, W., & Sugiyama, K. (1995). Layout adjustment and the mental map. *J. Visual Languages and Computing*, 6(2): 183-210.

Moody, J., McFarland, D., & Bender-deMoll, S. (2005). Dynamic Network Visualization. *American Journal of Sociology,* 110(4), 1206-1241.

Otte, E., & Rousseau, R. (2002). Social network analysis: a powerful strategy, also for the information sciences. *Journal of Information Science,* 28(6), 441-453.
Salton, G., & McGill, M. J. (1983). *Introduction to Modern Information Retrieval*. Auckland, etc.: McGraw-Hill.
Theil, H. (1972). *Statistical Decomposition Analysis*. Amsterdam/ London: North-Holland.
Tijssen, R., de Leeuw, J., & van Raan, A. F. J. (1987). Quasi-Correspondence Analysis on Square Scientometric Transaction Matrices. *Scientometrics* 11, 347-361.
Wasserman, S., & Faust, K. (1994). *Social Network Analysis: Methods and Applications*. New York, etc.: Cambridge University Press.
Zhou, P., & Leydesdorff, L. (2006). The emergence of China as a leading nation in science. *Research Policy,* 35(1), 83-104.